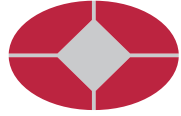

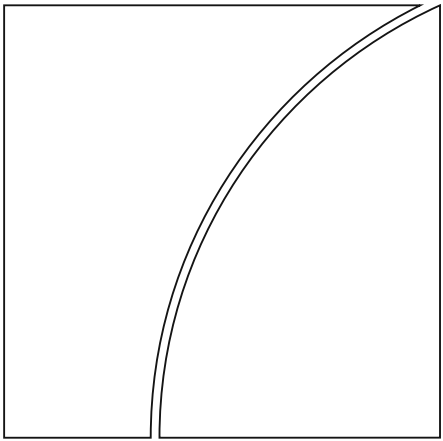

# BIS Working Papers
## No 980
## What does machine learning say about the drivers of inflation?

by Emanuel Kohlscheen

Monetary and Economic Department

November 2021 (this version December 2022)

# What does machine learning say about the drivers of inflation?

Emanuel Kohlscheen[1,2]

**Abstract**

This paper examines the drivers of CPI inflation through the lens of a simple, but computationally intensive machine learning technique. More specifically, it predicts inflation across 20 advanced countries between 2000 and 2021, relying on 1,000 regression trees that are constructed based on six key macroeconomic variables. This agnostic, purely data driven method delivers (relatively) good outcome prediction performance. Out of sample root mean square errors (RMSE) systematically beat even the in-sample benchmark econometric models. Partial effects of inflation expectations on CPI outcomes are also elicited in the paper. Overall, the results highlight the role of expectations for inflation outcomes in advanced economies, even though their importance appears to have declined somewhat during the last 10 years.

---

[1] Bank for International Settlements. Centralbahnplatz 2, 4002 Basel, Switzerland. *E-Mail address*: emanuel.kohlscheen@bis.org.
[2] I am grateful to Deniz Igan and Daniel Rees for providing useful comments. The views expressed in this paper are those of the author and do not necessarily reflect those of the Bank for International Settlements.

# 1. Introduction

What are the key drivers of inflation? And which role do expectations play in the inflation process? These have been long standing questions in macroeconomics, particularly given their high relevance to economic policy making. Indeed, the paper that is often credited with having started the rational expectations revolution (Muth (1961)) was concerned exactly with the above questions.

The current study attempts to shed some fresh light on these core macroeconomic questions. It does so through the lens of a flexible non-parametric data driven method. Specifically, it applies the well-established random forest approach (Breiman et al (1984), Breiman (2001)) to disentangle the drivers of inflation since 2000 across 20 advanced economies. Random forests are an ensemble of regression trees, each of them constructed based on simple and intuitive sequences of sample partitions. Besides their attested good prediction performance (see e.g. Fernandez-Delgado et al (2014), Chakraborty and Joseph (2017) and Medeiros et al (2021)), this method has the advantage that "great out-of-sample performance [comes] without requiring subtle tuning" (Athey and Imbens (2019), p. 696). On top of comparing explanatory performance with traditional econometric benchmarks, as far as possible, this article also tries to interpret the economic reasons that are behind the (relative) success of the technique in explaining recent consumer price inflation.

Overall, the analysis attests the relative strong performance of the random forest model in predicting contemporaneous and future headline and core CPI inflation, even when only a small standard set of macroeconomic indicators is used. In fact, the out of sample root mean square error (RMSE) of the machine learning (ML) model beats even the in-sample performance of standard OLS using the same set of explanatory variables/features – which are firmly grounded on economic theory. This suggests that non-linearities play an important role in explaining inflation.

Overall, expectations emerge as the most important predictor of CPI inflation, followed by past inflation. That said, the importance of expectations has declined during the last 10 years. During this period, the partial effects that are teased out from the random forest model point to a flattening of the effects of expectations when these are above 2%. Throughout, exchange rate variations are found to add relatively little value in predicting inflation outcomes.

**Relation to the literature.** The study builds on a growing literature that applies machine learning (ML) to economics. Kleinberg et al (2015) discuss the advantages and caveats of applying ML techniques to economic prediction problems. They argue that ML provides a disciplined non-parametric way to predict economic outcomes. Mullainathan and Spiess (2017) offer an example of how regression trees can be used to better predict house prices. They conclude their review by stating that "*machine learning provides a powerful tool to hear, more clearly than ever, what the data have to say*". As such, it can be a useful complement to more traditional model based methods.[3]

Chakraborty and Joseph (2017) compare the inflation prediction performance of 10 econometric and machine learning models for the United Kingdom. They find that, post-GFC, random forests provide the best prediction performance among stand-alone models in the testing sample.[4] More recently, Medeiros et al (2021) compare different ML methods' performance to predict inflation in the United States. Also, these authors conclude that random forests dominate all other methods. These findings confirm more general ones by Fernandez-Delgado et al (2014) – who compared the performance of 179 classifier models across 121 datasets and found (the relatively simple) random forest to be the top performer among all options. Coulombe (2021) provides an early attempt to combine the random forest methodology with a macroeconomic model.

The current paper contributes to deepen our understanding of the drivers of inflation relying on a purely data driven method. The results reassert the importance of expectations in the price formation process. This emerges from the relative inflation predictor importances and from eliciting the partial effects of expectations on inflation outcomes. In this sense, it strengthens the case for current inflation targeting frameworks.[5] Second, it highlights the importance of non-linearities, for instance for the effect of the output gap on inflation, as well as for how expectations translate into price pressures. Third, it finds that oil price movements and global PPI inflation are also important drivers of CPI inflation in advanced economies, indicating a global dimension of inflation – particularly after 2010. In this respect, the paper is also related to the large literature that examines the role of global drivers of inflation.[6] Interestingly, a recent paper by Kamber and Wong (2020) finds that while global factors do play a substantial role in explaining the inflation gap, they generally do not explain the inflation trend in advanced economies – suggesting that policies do play a key role.[7]

**Outline.** The article proceeds as follows. Section 2 explains the methodology and the dataset that was used to predict current and future inflation. Section 3 presents the baseline results. Section 4 explores how drivers of inflation have been changing over time. Section 5 examines the performance of random forests in forecasting inflation 6 and 12 months ahead. Section 6 presents several

---

[3] Earlier, Varian (2014) provided an example of regression trees for predicting mortgage approvals.
[4] Page 56, Table 8.
[5] The overall importance of expectations for inflation was recently called into question, following an article by Rudd (2021).
[6] See e.g. Borio and Filardo (2007), Monacelli and Sala (2009), Cicccarelli and Mojon (2010), Neely and Rapach (2011), Mumtaz and Surico (2012), Gillitzer and McCarthy (2019) and Forbes (2019), among others.
[7] Note that, as in Jasova et al (2019) and Forbes (2019), in this paper policies are captured indirectly through their effects on expectations, past inflation, output gaps and exchange rates.

robustness checks. The paper concludes by suggesting avenues for further research.

## 2. Modelling Headline Inflation with Regression Trees

### 2.1. Data and feature selection

Given that most central banks calibrate their monetary policy based on targets for headline CPI inflation, this study assesses the drivers of this indicator for a broad set of advanced economies. More specifically, it compares the performance of standard regression techniques with that of a well established machine learning technique (random forests) in predicting contemporaneous and future quarterly seasonally adjusted CPI inflation in 20 advanced countries (the dependent variable).[8] The choice of random forests is based on earlier results obtained by Fernandez-Delgado et al (2014), Chakraborty and Joseph (2017) and Medeiros et al (2021).

The countries that are included in the study are all that have a population of at least one million and current GDP per capita above $25,000. They are Austria, Belgium, Canada, Czechia, Denmark, Estonia, Finland, France, Germany, Ireland, Italy, the Netherlands, Norway, Portugal, Slovenia, Spain, Sweden, Switzerland, the United Kingdom and the United States. The time span of the analysed data goes from 2000 to mid-2021.

As potential explanatory variables, key factors that have been found to be of importance in the theoretical and applied literature are included. Broadly speaking, the list of factors is fairly similar to that used in recent papers by Forbes (2019) and Jasova et al (2019), among others. They include six variables: lagged inflation, to capture the persistence of the inflation process; 12-month ahead inflation expectations, as surveyed on a monthly basis by *Consensus Economics* from a representative group of banks;[9] the output gap, which was computed after applying a one-sided HP filter to the real GDP series;[10] the cumulative percentage variation of the oil price (Brent) over a year; the similar variation of the BIS nominal effective exchange rate for each country and the average producer price inflation (PPI) measure in the three major economies (the United States, the euro area and China). The latter captures factory gate inflation, and is likely to reflect the impact of input costs. Together with oil prices, it also captures an international dimension of CPI inflation.[11] [12] Note that by limiting the set of explanatory variables ("features") that are used to those that are firmly grounded in the literature, the advice of Chakraborty and Joseph (2017) to limit complexity from the onset is followed.

---

[8] Obtained directly, in seasonally adjusted form, from the OECD.
[9] Carrol (2003) finds that the dynamics of expectations are well captured by a model in which households' expectations derive from news reports about professional forecasters' expectations.
[10] Contrary to the two-sided version, the one-sided version of the filter does not rely on information after the quarter in question.
[11] Labour market slack is included indirectly through the output gap.
[12] Data are sourced from the BIS, Consensus Economics, the OECD and Bloomberg.

The average annualized headline inflation in the panel is 1.87%, with a standard deviation of 3.20%. For reference, average core inflation is 1.59% (standard deviation of 1.61%). 12-month ahead headline inflation expectations average 1.80%, with a standard deviation of 0.89%.[13]

Pooled OLS based on the six explanatory variables is able to explain 39% of the variation in contemporaneous quarter headline inflation outcomes (see Table A1 in the Appendix). All six correlates clearly correlate strongly with inflation, with robust *t*-statistics that vary between 2.0 (for the output gap) and 42.1 (for inflation expectations). The *F*-statistic of the model is 523.0 (*p-value* < 0.001).

## 2.2. Growing the Random Forests

CPI inflation is predicted by means of regression trees and random forests, as outlined in Breiman *et al* (1984) and Breiman (2001). The main advantage of these methods is that they are able to accommodate non-linearities, as well as to capture potentially complex interactions between the explanatory variables. Essentially, inflation prediction is treated as a classification problem.[14] While highly flexible as to what concerns outcomes, the algorithm follows a highly disciplined and transparent logic.

The regression tree algorithm mechanically grows trees based on successive splits of the panel according to an explanatory variable (the feature) and an associated threshold level that minimizes the sum of squared residuals after the split. This is done again and again for each new node until a pre-defined tree depth is reached (the "stopping criterion").[15]

To put it in mathematical terms, the splitting of observations at each node is based on a choice of predictor $k$ and an associated threshold $t_k$ at which the split happens, selected according to the expression

$$arg\ min\ (k, tk)\ \frac{(S_A \cdot MSE(S_A) + S_B \cdot MSE(S_B))}{S},$$

where $S$ corresponds to the number of training points, and $A$ and $B$ represent the subsets of the training sample after a binary split.

The algorithm then takes the average of the target variable over all observations in a given final node as a prediction, $((1/n_{\text{obs in leaf}}) \bullet \Sigma_{i\ \epsilon\ \text{leaf}}\ y_i)$, and examines it's difference with actual values of the variable of interest, $y_i$, for each observation. The end objective is to minimize the sum of squared residuals over all the final nodes (leaves).

---

[13] These were computed based on a properly weighted average of current and next year forecasts. That is, with geometric weights that are based on the number of the next 12 months that fall within the current year and the following year.
[14] Note that the analysis first looks at contemporaneous prediction, i.e. model fit in the econometric sense. Later we analyse actual forecasts.
[15] Put differently, the method is based on exhaustive recursive partitioning of the sample.

To introduce randomness, a large ensemble of trees is built, based on randomly selected subsamples. For training each tree, the algorithm keeps one third of the sample out-of-bag for posterior testing and grows regression trees based on the remaining observations (the "training sample"). Instead of relying on only one tree, outcome predictions are then based on the average values of the outcome variable for terminal nodes of a large number of trees, that is the *random forest*. At the same time, to introduce further randomness, only a random subset of the features is utilized at each split. Such randomness is meant to increase robustness of the predictions for out-of-sample data.

The main advantage of using a large number of trees is that the variance of the predictions declines and overfitting is minimized (Friedman et al (2009), Chakraborty and Joseph (2017)).[16] Figure 1 shows that, for the current application, the overall MSE declines very rapidly as the number of trees grows. Most gains in accuracy are already evident after the threshold of 10 trees is passed.

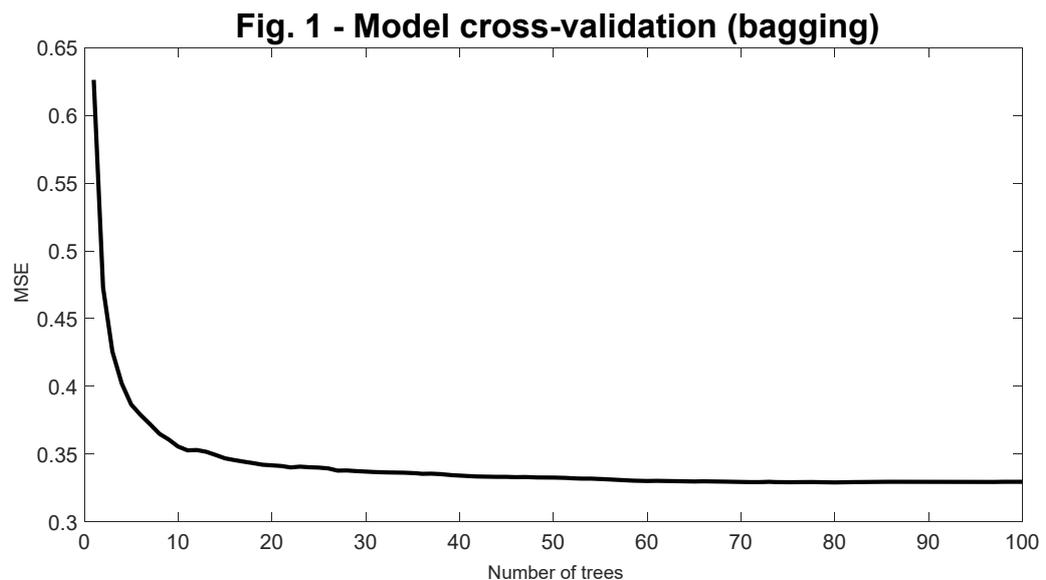

### 2.3. Measuring Performance

Table 1 shows model RMSEs for inflation, as well as RMSE ratios between the regression tree method and traditional econometric benchmarks for forests of 100 and 1,000 trees. It does so for different pre-specified tree depths. The parameter that controls the depth of grown trees (pruning) is the minimum number of observations per parent node (which is denoted *p*). A lower minimum parent size of a splitting node means that deeper trees are grown. Too deep trees generally imply overfitting, which may come at the cost of reducing its flexibility and out of sample performance. Shallower trees can be more robust for use with new incoming data.

---

[16] Mentch and Zhou (2020) show that "*the additional randomness injected into individual trees serves as a form of implicit regularization, making random forests an ideal model in low signal to noise (SNR) settings.*"

The asymptotic limit to gains in predictive performance that is apparent from Graph 1 is again reflected in the fact that RMSEs are only marginally lower when an ensemble of 1,000 trees is used instead of 100 trees (–0.9% for $p$=10, out of sample). For the inflation panel in question, varying the minimum parent node size between 4 and 30 has negligible effects on out of sample prediction performance. That is, the cost of overfitting in this application is found to be minimal. For this reason, a minimum of 10 observations per splitting node is used as the benchmark throughout. Later, it is shown that conclusions change very little when a minimum parent size of 30 is used instead.[17]

**Table 1 – Absolute and Relative RMSEs for CPI inflation**

|  | ML RMSEs | | ML RMSEs relative to in sample | | ML RMSEs relative to out of sample | |
|---|---|---|---|---|---|---|
| min obs/splitting node | in sample | out of sample | | | | |
| with 100 regression trees | | | AR(1) | OLS | AR(1) | OLS |
| 4 | 0.279 | 0.573 | 0.352 | 0.452 | 0.734 | 0.929 |
| 6 | 0.299 | 0.570 | 0.377 | 0.484 | 0.731 | 0.925 |
| 8 | 0.319 | 0.571 | 0.402 | 0.517 | 0.733 | 0.928 |
| 10 | 0.340 | 0.572 | 0.428 | 0.550 | 0.733 | 0.928 |
| 12 | 0.355 | 0.574 | 0.447 | 0.575 | 0.736 | 0.931 |
| 14 | 0.370 | 0.574 | 0.466 | 0.598 | 0.736 | 0.932 |
| 16 | 0.381 | 0.572 | 0.480 | 0.616 | 0.733 | 0.929 |
| 18 | 0.394 | 0.576 | 0.496 | 0.637 | 0.738 | 0.935 |
| 20 | 0.404 | 0.577 | 0.509 | 0.654 | 0.740 | 0.937 |
| 30 | 0.440 | 0.577 | 0.554 | 0.712 | 0.739 | 0.936 |
| 60 | 0.499 | 0.589 | 0.628 | 0.807 | 0.755 | 0.956 |
| 120 | 0.548 | 0.604 | 0.690 | 0.887 | 0.774 | 0.980 |
| with 1,000 regression trees | | | | | | |
| 4 | 0.276 | 0.565 | 0.347 | 0.446 | 0.724 | 0.917 |
| 6 | 0.297 | 0.565 | 0.374 | 0.480 | 0.724 | 0.917 |
| 8 | 0.319 | 0.567 | 0.401 | 0.516 | 0.727 | 0.920 |
| 10 | 0.337 | 0.567 | 0.424 | 0.545 | 0.727 | 0.920 |
| 12 | 0.353 | 0.567 | 0.445 | 0.571 | 0.726 | 0.920 |
| 14 | 0.368 | 0.569 | 0.463 | 0.595 | 0.729 | 0.923 |
| 16 | 0.381 | 0.569 | 0.479 | 0.616 | 0.730 | 0.924 |
| 18 | 0.392 | 0.570 | 0.494 | 0.634 | 0.730 | 0.925 |
| 20 | 0.402 | 0.571 | 0.506 | 0.650 | 0.731 | 0.926 |
| 30 | 0.440 | 0.575 | 0.554 | 0.711 | 0.737 | 0.933 |
| 60 | 0.499 | 0.586 | 0.628 | 0.807 | 0.751 | 0.951 |
| 120 | 0.547 | 0.601 | 0.689 | 0.885 | 0.770 | 0.975 |

Note: Ratios below indicate that ML outperforms AR(1) or OLS models.

---

[17] The total number of observations is 4,941. Note that even though quarterly annualized inflation is the target variable, the fact that the variables are available at monthly frequency for the 20 countries in question is used. That is, for each month the quarterly variation is computed by comparing indicators in month $t$ and $t$-3.

Strong relative performance in predicting current quarter inflation is clear. RMSE ratios are below 1 throughout (i.e. for all the 96 simulations), indicating consistently superior performance of machine learning relative to AR(1) and OLS models. As it turns out, the out of sample RMSE of 0.567 for the benchmark tree depth ($p$ = 10, $n$ = 1,000) beats even the in-sample RMSEs of econometric benchmarks (absolute RMSEs of, respectively, 0.794 and 0.618).[18] That is the reduction in RMSEs is of 28% and 8%.

Comparing likes with likes, machine learning delivers a RMSE reduction of 58% relative to the AR(1) model and of 46% relative to OLS in sample. Of course, part of this is because of its much higher capability to fit the data, when compared to linear methods that disregard interactions. That said, the magnitude of the reduction may be taken to suggest that non-linearities and interactions may be an important aspect to consider when analysing the data. The machine learning method thus provides a powerful method for describing the contemporaneous patterns of inflation. Out of sample, the reduction in RMSEs is of 27% relative to the naïve AR(1) benchmark, and of 8% relative to OLS.

## 3. Examining Drivers of CPI inflation

Which factors have been the key drivers of CPI inflation in advanced economies since the turn of the millennium? The ML answer to this fundamental question can be obtained by looking at relative predictor importance in the benchmark tree ensemble (i.e., $p$=10, $n$=1,000). This statistic summarizes the contribution of each factor for the reduction in RMSEs over all splits and trees. Figure 2 shows that, on the whole, inflation expectations come out on top, followed closely by past inflation, which captures inertia in price setting. These two factors are followed by oil price changes and global PPI inflation. Exchange rate movements matter less, which is in line with earlier evidence of reductions in pass-throughs (e.g. Jasova et al (2019)).

---

[18] The out of sample RMSEs for econometric models was obtained by running the estimates on 2/3 of the sample, and projecting the outcome for the remaining 1/3 of the sample several times.

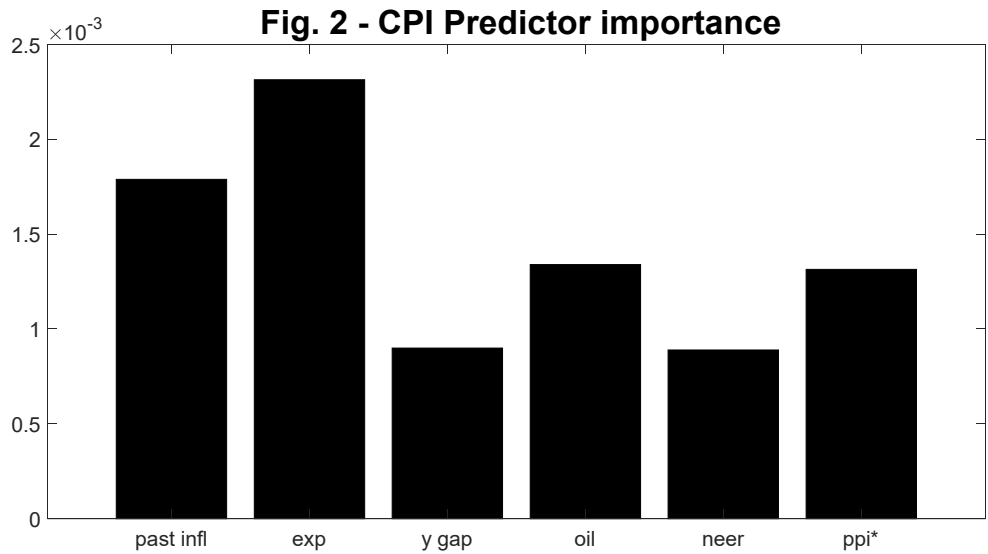

The partial effects of expectations on actual inflation can be inferred by looking at predicted headline inflation for different levels of 12-month ahead inflation expectations. This is done while all other variables are held fixed at their mean levels. This offers an interpretation as those of regression coefficients in econometrics, with the key difference that effects do not need to be linear. They also do not need to conform to any pre-defined form. At the same time, such evidence should be interpreted as being only suggestive, as it is not based on a structured econometric theory, but instead offers a data description. With this caveat in mind, the outcome of such exercise for *p*=10 and *n*=1,000 is plotted in Figure 3.

Higher inflation expectations are clearly associated with higher inflation outcomes. The average slope over the interval shown is 1.1. The effects of expectations appear larger at intermediate levels of this variable, while they tend to diminish when expectations are very low/high. At the mean of other covariates, when inflation expectations are exactly at 2%, the predicted outcome is 1.85%. In turn, when expectations are at 4%, predicted inflation is 1.10 percentage point above that.

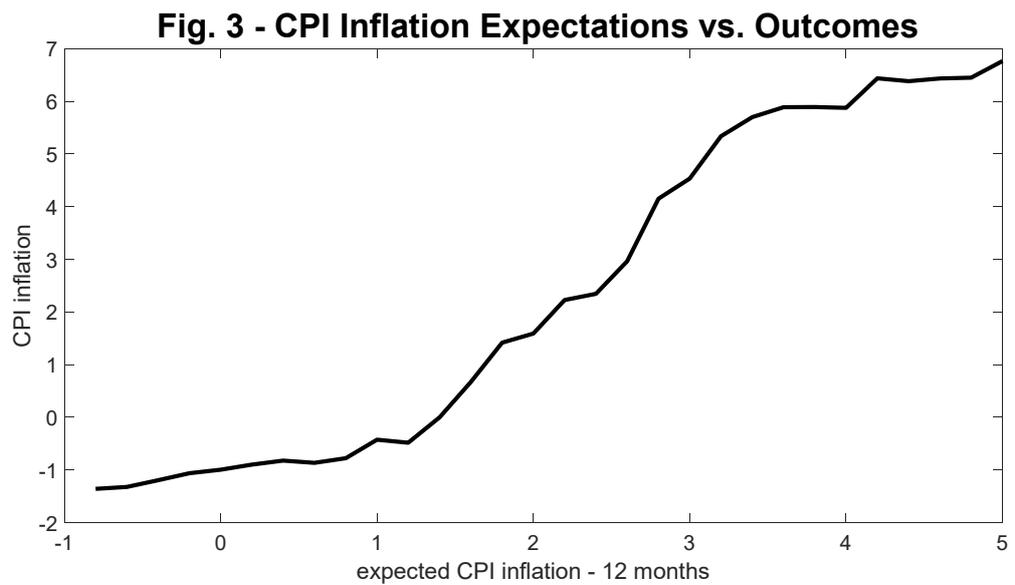

Fig. 3 - CPI Inflation Expectations vs. Outcomes

A similar analysis can be done for the output gap, which is a factor that features prominently in most theoretical models of inflation and monetary policy reaction functions. The simulated partial effects confirm the positive association with inflation levels across the 20 countries (Figure 4). What is noteworthy is that the effects are far from linear, as most of the link manifests itself when the output gap is below −1. That is, economic activity that is far below potential (as estimated by a one-sided HP filter) has a strong dampening effect on inflation. On the other hand, economic overheating does not seem to boost inflation – at least insofar other factors such as oil prices and the exchange rate are held constant. This is of course consistent with modest inflation pre-Covid, despite robust activity readings and historically low unemployment rates in many countries.[19]

---

[19] In the current context, these results would appear to suggest that the post-2021 surge in CPI inflation across the world is most likely due to severe supply disruptions, rather than excess demand.

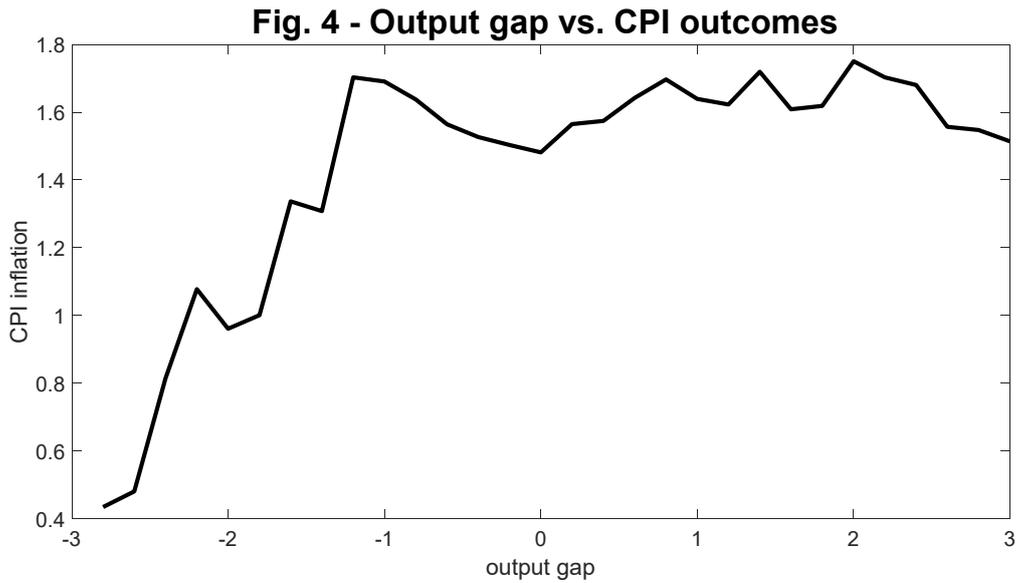

## 4. Time Variation

To explore how the role of drivers of inflation may have changed over time, the panel is split according to decade. What this exercise reveals is that during the last decade, the role of expectations has diminished somewhat, while oil price variation has become a more prominent driving feature (Figure 5). In both decades exchange rate variations were the least relevant for predicting inflation.

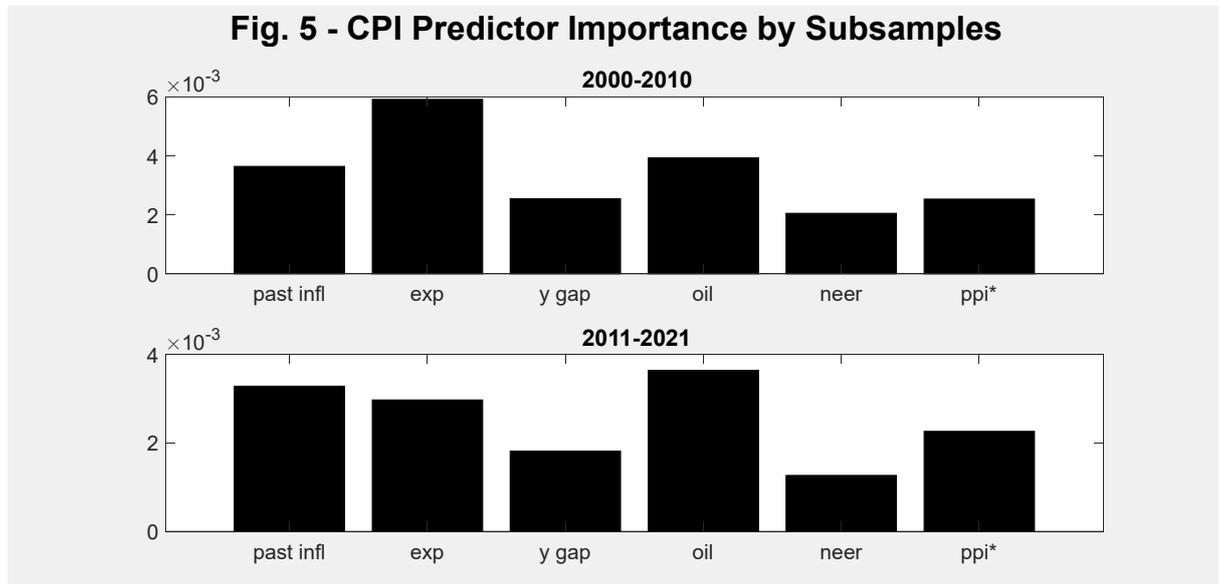

Examining the partial effects also confirms the more subdued effect of 12-month ahead inflation expectations on CPI outcomes (Figure 6). The average slope almost halves from 1.37 in 2000-2010, to just 0.75 in the last 10 years.

Interestingly, the effect of higher expected price momentum on inflation appears to flatten at much lower thresholds since 2011, relative to the whole sample case.

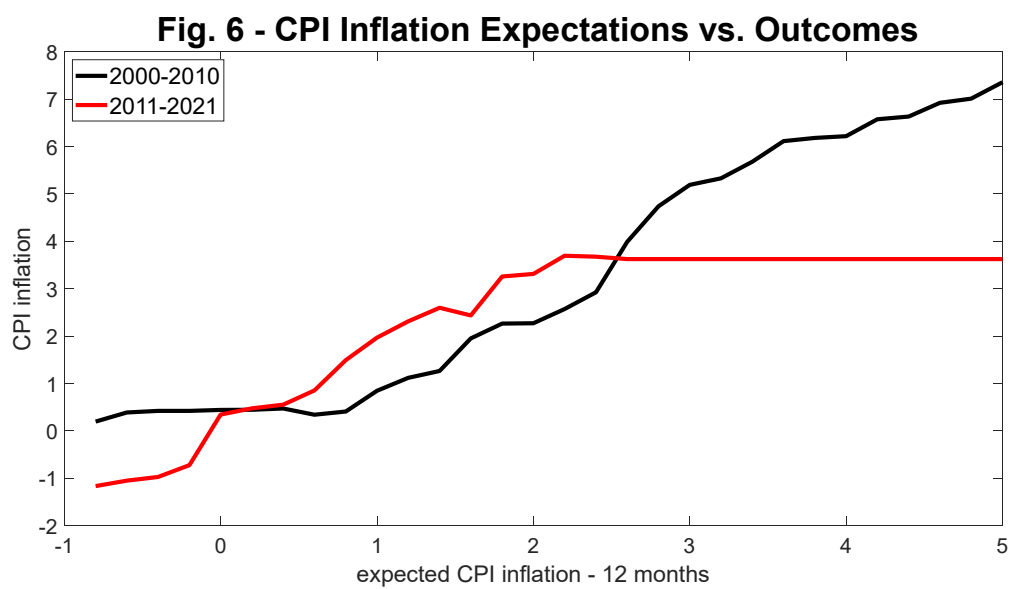

## 5. Forecasting 6 and 12 Months Ahead Inflation

The previous sections focused on identifying the key drivers of inflation, rather than forecasting inflation ahead of time. As such, they examined current quarter inflation. Yet broadly similar results obtain when the same information set is used to forecast CPI inflation 6 or 12 months down the road. Table 2 shows that RMSE ratios are again all below unity, with RMSE error reductions between 10% and 14% relative to OLS.[20]

**Table 2 – Relative Performance in Forecasting CPI inflation 6 and 12 Months Ahead**

|  | ML RMSEs | | ML RMSEs relative to out of sample | |
| --- | --- | --- | --- | --- |
|  | in sample | out of sample | AR(1) | OLS |
| forecast horizon | | | | |
| 6 months | 0.329 | 0.553 | 0.764 | 0.904 |
| 12 months | 0.260 | 0.429 | 0.738 | 0.860 |

Note: Ratios below indicate that ML outperforms AR(1) or OLS models.

---

[20] Even larger RMSE reductions can be obtained if in a first step one includes also country dummies as features, and then eliminates the three or five countries whose dummies have the largest predictor importance factor (i.e. the most heterogeneous countries with regards to inflation). In this paper, the choice was not to include country dummies in the baseline specification, as doing so would make predictions country specific, making a general assessment more difficult. Obviously, this cost may well be worthwhile for practical forecasting exercises.

As for the relative importance of the predictors, expectations continue to be the key driver for 12-month forward inflation. Global PPI inflation becomes the second most relevant factor (Figure 7). Figure 8 shows how predicted inflation covaries with these variables.

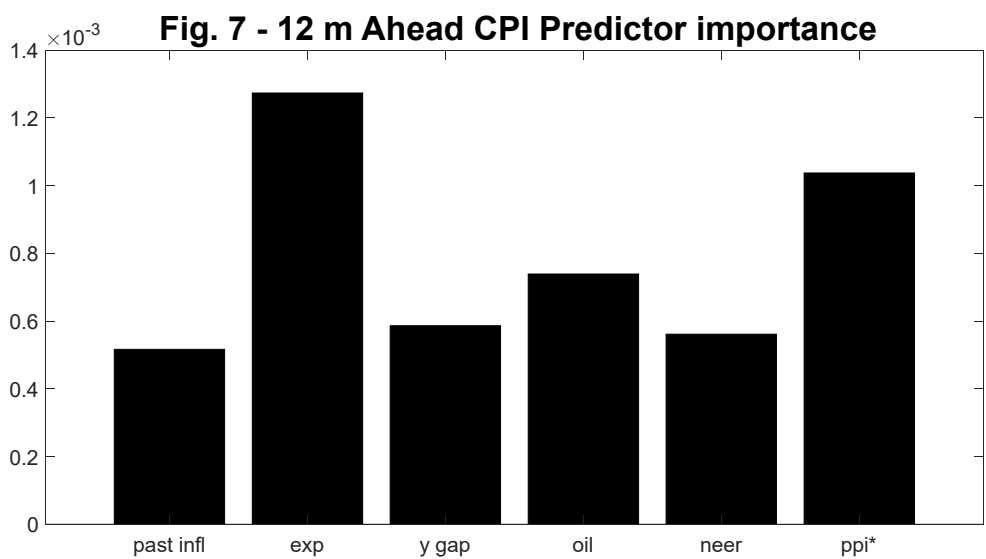

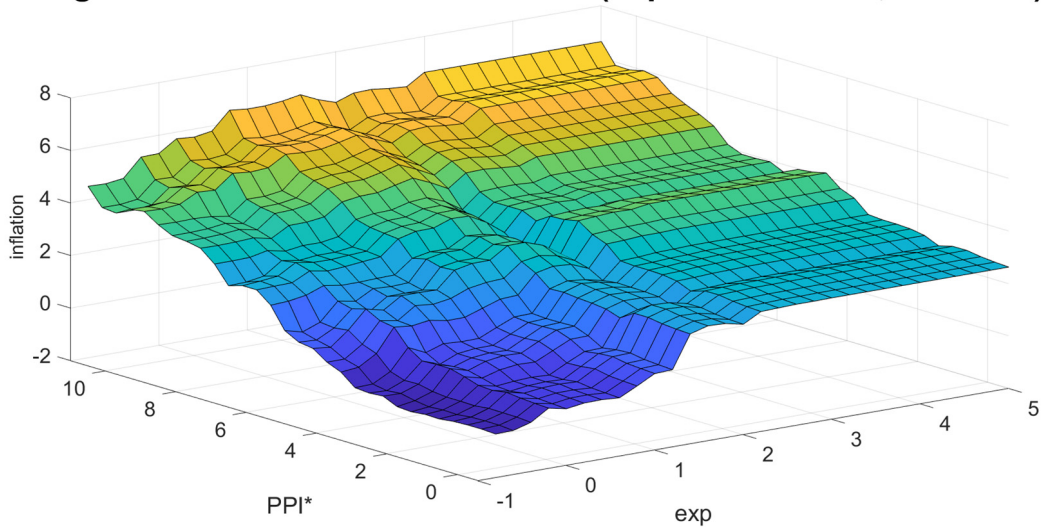

## 6. Alternative Tree Depths and Core Inflation Prediction

Several robustness checks of the above findings were carried out. On the whole, they confirmed the above conclusions. First, oil price and exchange rate variation were taken for a single quarter, instead of the cumulative effects over 12 months. This switch did not produce noticeable changes.

Second, the simulations were repeated using much shallower regression trees. Figures 9 and 10 show that tripling the minimum number of required observations for splitting nodes produced little changes.

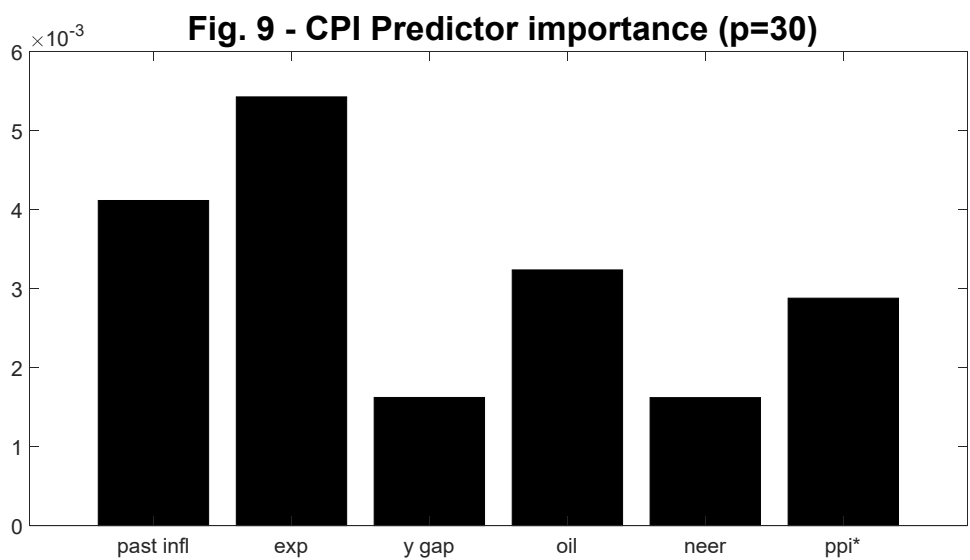

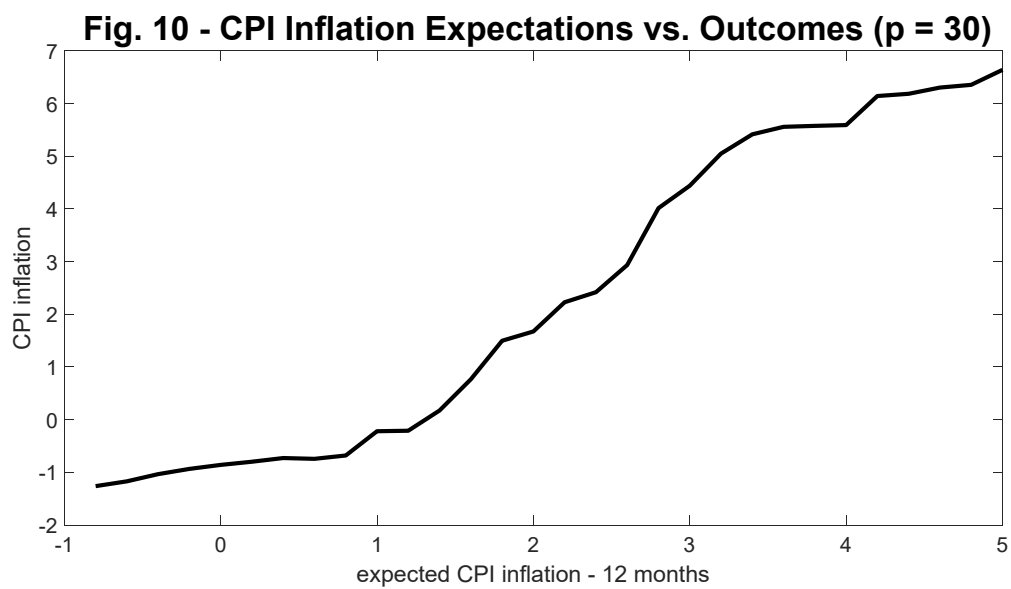

Third, the target variable (and lagged inflation) was replaced by the core CPI inflation rate (that is, based on the OECD CPI index that excludes food and energy). A fundamental caveat to bear in mind for such exercise is that expectations for core inflation are not available. Even so, inflation expectations do come out as the most important driver of core inflation as well (Figure 11). Naturally, as it has been stripped out of the index, the importance of oil declines to a level that is below that of the output gap. The remaining effects are likely due to second round effects of oil price variations on the general price index. Further, the partial effects of expectations on core CPI inflation turn out to be broadly similar to those of headline inflation (Figure 12).

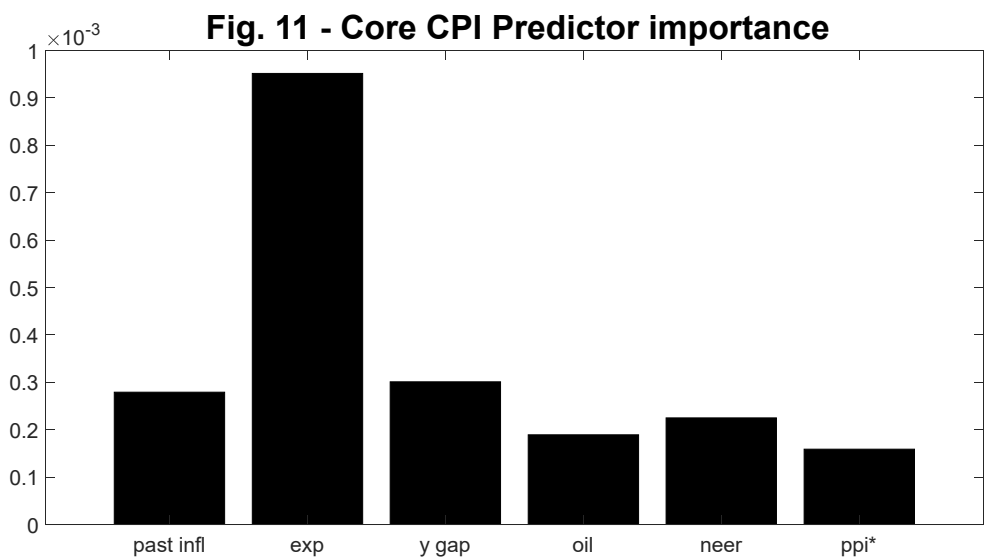

Fig. 11 - Core CPI Predictor importance

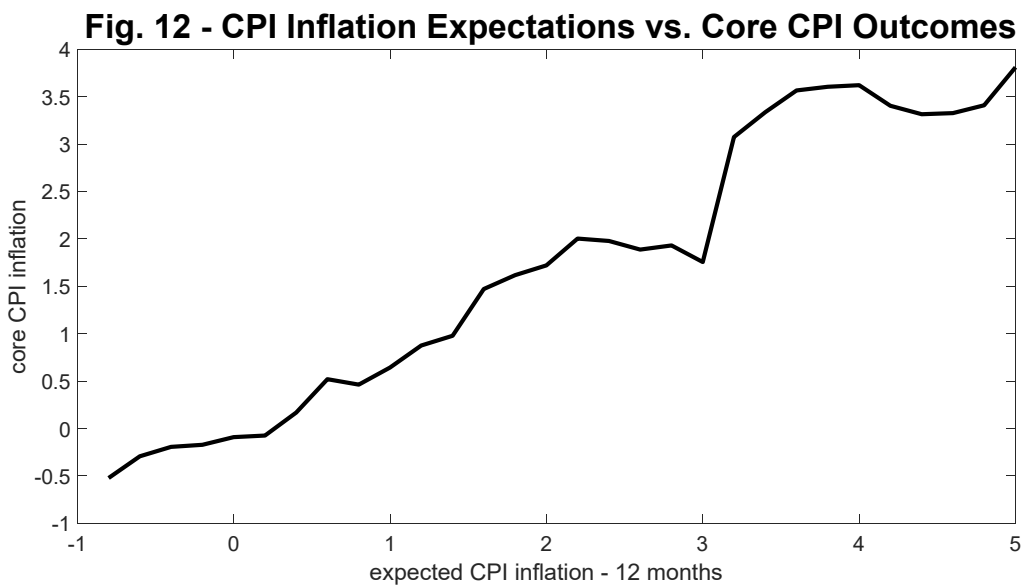

Fig. 12 - CPI Inflation Expectations vs. Core CPI Outcomes

## 7.   Concluding Remarks

All in all, the analysis confirms the usefulness of random forests for explaining inflation patterns in a broad set of advanced economies. The method delivers much better in-sample description, as well as better inflation forecasts. In the panel sample of this paper, RMSE reductions for 6- to 12-month ahead inflation forecasts relative to simple linear methods lie between 10% and 14%. This suggest that this disciplined and intuitive computational method is able to better capture non-linearities and possibly complex interactions between factors that occur in the real world.

Importantly, the findings concerning the drivers of inflation are based on a limited set of explanatory variables – all of which are firmly grounded in the theoretical and empirical literature. They highlight the key role of expectations for inflation, even though their relative importance in advanced economies appears to have declined recently.

Future research should aim to further enhance the economic interpretation of ML tools. The current paper made a first step in this direction by eliciting the partial effects of key explanatory variables. Another potentially fruitful avenue for future research is the combination of powerful machine learning techniques with macroeconomic models.

**Drivers of CPI inflation** Table A1

| Dependent variable: Headline CPI inflation (quarterly, s.a. annualised) | | |
|---|---|---|
| | AR(1) | Benchmark |
| constant | 2.113*** | –1.171*** |
| *t-stat* | 41.26 | 13.98 |
| lagged inflation | –0.130*** | –0.458*** |
| *t-stat* | 9.38 | 35.49 |
| 12-month ahead inflation expectation | | 1.989*** |
| *t-stat* | | 42.11 |
| output gap | | 0.032** |
| *t-stat* | | 2.02 |
| Brent oil price increase | | 0.011*** |
| *t-stat* | | 9.11 |
| nominal effective e.r. appreciation | | –0.066*** |
| *t-stat* | | 7.47 |
| PPI inflation in 3 major economies | | 0.251*** |
| *t-stat* | | 14.60 |
| observations | 4,941 | 4,941 |
| R2 | 0.017 | 0.389 |
| F statistic | 88.0*** | 523.0*** |

Note: Time span is from 2020Q1 to 2021Q2. Reported *t*-statistics below coefficients are based on robust standard errors. ***/**/* denote statistical significance at 1, 5 and 10%, respectively.

## Annex Table – Summary statistics

| | mean | std. dev. |
|---|---|---|
| CPI inflation | 1.87 | 3.20 |
| global PPI inflation | 1.56 | 3.61 |
| output gap (1-sided HP) | -0.30 | 2.27 |
| 12 m inflation expectation | 1.80 | 0.89 |
| nom. eff. exchange rate change | 0.50 | 4.08 |
| Brent price change | 12.15 | 39.35 |

# Previous volumes in this series

All volumes are available on our website www.bis.org.